# Focused Ion Beam patterning of MnSb(1-101) based spintronic devices

S. N. Holmes*[1], J. Gough[2], E. Dommett[3], G. R. Bell[3]

[1]Department of Electronic and Electrical Engineering, University College London, Torrington Place, London, WC1E 7JE, United Kingdom

[2]London Centre for Nanotechnology, University College London, 17-19 Gordon Street, London, WC1H 0AH, United Kingdom

[3]Department of Physics, University of Warwick, Coventry CV4 7AL, United Kingdom

**Abstract**

Low temperature transport measurements are presented of ferromagnetic MnSb(1-101) devices with the magnetic properties patterned using a $Ga^+$ focused ion beam-FIB system at 30 keV. FIB patterning introduces disorder and this is quantified in this paper through measurements of the longitudinal resistivity, $\rho_{xx}$ and the anomalous Hall effect contribution to $\rho_{xy}$. The MnSb(1-101) structural phase is the niccolite phase with a surface state in addition to bulk states. FIB doses up to $\sim 1\times10^{16}$ $Ga^+$ ions/cm$^2$ reduces the anisotropic magnetoresistance signal but increases the size of the anomalous Hall effect signal in out-of-plane magnetic fields, B. The anomalous Hall effect contribution to the conductivity $\sigma_{xy}$ is $\sim e^2/h$ in the undamaged devices, where e is the electronic charge and h is Planck's constant. This quantity is sensitive to the level of disorder induced by the $Ga^+$ ion beam, reducing to zero at dose levels $>1\times10^{16}$ $Ga^+$ ions/cm$^2$. $\rho_{xx}$ shows a log(B) dependence after the magnetization has saturated with the low field anisotropic magnetoresistance contribution $\sim 0.12\%$. The conductivity change ($\Delta\sigma_{xx}$) is $\sim e^2/h$ in the magnetic field range of log(B) behavior. $\rho_{xx}$ shows a reduced fit to a log(B) dependence at high FIB dose levels and is dependent on the damage uniformity. Patterning nanostructured magnetic behavior in MnSb, with compatibility to altermagnetic materials, in particular the niccolite phase of CrSb and non-trivial Berry phase contributions to transport make this ferromagnetic material and patterning technique useful for future spintronic device development.

*corresponding author e-mail: stuart.holmes@ucl.ac.uk

# I. INTRODUCTION

The integration of ferromagnetic MnSb with reduced dimensional devices fabricated in InSb and InAs-based structures is now possible with spin-injection into these high spin-orbit systems an attractive proposition for the interrogation of topological phases or added functionality such as combining altermagnetic materials and devices. Growth details are important, for example in the cubic phase MnSb can be half-metallic with surface properties different from those of the bulk [1]. Can MnSb devices be patterned using a focused ion beam (FIB) and does the predominant niccolite phase of MnSb/GaAs have a surface state that can be patterned? In-situ processing could be envisaged then with further MBE (Molecular Beam Epitaxy) growth, either of capping layers or carrier confinement layers. Patterning of the magnetic properties can be on the sub-10 nm scale, limited by the resolution of the FIB system.

The effects of $Ga^+$ ion damage on the magnetic properties of MnSb are presented in this paper by measuring resistivity components $\rho_{xx}$ and $\rho_{xy}$ in applied magnetic fields at low temperature. The anomalous Hall effect (AHE) in $\rho_{xy}$ is dominant in this material and this is reported in detail here. In fact, earlier work [2] identified an influence from a Berry curvature on $\rho_{xy}$ and a spin re-orientation transition at 100 K in MnSb but with no detail on crystal structure, rather bulk film properties. The crystalline structure details are important in MBE grown MnSb both in terms of magnetic anisotropy behavior and the possible manifestation of surface states [3]. MnSb(0001) (the niccolite phase that is studied here) stabilized on a GaAs(111) substrate can be used as a buffer layer for the altermagnetic material CrSb (0001) [4] due to the close lattice match required for coherent epitaxy. In Ref. [4] it had not yet been established that CrSb has an altermagnetic state, in addition to the established antiferromagnetic ordering [5].

The cubic (c) phase of MnSb is predicted to be a room temperature half-metal [1] although this phase is generally unstable in MBE growth. MBE growth results in predominantly the niccolite phase on GaAs, InGaAs, InGaSb III-V substrates with the c-phase only stable for small grain sizes [6, 7]. There are recent reports on related materials and devices, e.g. PtMnSb which is a Heusler alloy material [8] with other phases or alloys of MnSb that result in either half metallic properties or topological insulator behavior for the case of $SrMnSb_2$ [9].

The MnSb surface state can dominate the electrical properties with Sb-terminated surfaces showing higher anisotropic magnetoresistance (AMR) but lower AHE signals [10]. As FIB disorder is introduced there is a correlation of electrical conductivity and magnetic behavior, with this type of ferromagnetic/paramagnetic on-set discussed from the point of view of a percolation model prediction [11] in thin films. The mechanisms of weak localization become relevant for studying the magnetic properties of FIB patterned MnSb. At FIB doses ~ $10^{16}$ Ga$^+$ ions/cm$^2$ the electrical signals that are indicators of ferromagnetic behavior have been quenched and this sets a lower dose limit for patterning MnSb without inducing an amorphous crystallinity.

In this paper we present the properties of FIB damaged MnSb devices and address the issue of whether the magnetic properties of MnSb can be patterned controllably using a FIB system. Section II is a summary of experimental methods such as MBE growth, device fabrication and FIB damage information. In Sec. III the details of the electrical measurements are described at low temperatures in applied magnetic fields. Section IV is a discussion of the electrical and magnetic properties and Sec. V is the summary and conclusions of the work.

## II. EXPERIMENTAL METHODS
### A. MBE growth

The MnSb wafers were grown by MBE on semi-insulating GaAs(001) substrates. This substrate orientation and growth mode results in the stable niccolite phase with crystal orientation MnSb(1-101) [12]. In-situ RHEED (Reflection High Energy Electron Diffraction) measurements confirm the symmetry of the starting GaAs surface and the development of the epitaxial MnSb surface. The total MnSb thickness is ~ 50 nm with mesa isolation after etching to 50 nm depth. A thin surface capping layer of GaOx (or GaSb) see Table [1], was incorporated to protect the surface MnSb from oxidation.

| wafer | substrate | cap layer | growth information |
|---|---|---|---|
| A | GaAs(001) | GaOx | RHEED signal |
| B | GaAs(001) | GaSb | RHEED signal |

**Table [1].** Wafer summary of the MnSb-based devices.

The MBE growth mode of MnSb can be endotaxial in this material combination [13] although there is no evidence of this in these wafers. The wafers are undoped layers but there is a natural p-type behavior in the MnSb, with carrier density similar to those reported in Ref. [10] ~ $10^{22}$ cm$^{-3}$, the normal Hall constant R$_H$ is < 0.04 Ω/T at 1.5 K. This restricts the devices to p-type structures and does limit the utilization in spintronic devices [14].

Alternative buffer layers can be used to stabilize MnSb, for example InGaSb. In this case the MnSb(10-10) plane is aligned with the InGaSb (001) surface and the in-plane [0001] direction in MnSb is parallel to the [1-10] direction in the InGaSb [7]. The magnetic uniaxial anisotropic, easy axis (EA) direction is parallel to [1-10] in the InGaSb the same direction for MnSb grown on GaAs (001). The magnetic anisotropy can also be thickness dependent in MnSb [15].

**B. Device processing**

Standard Hall bar structures were fabricated using optical lithography with an HCl:H$_2$0 (1:10) mesa etch at 20 ºC. The length-to-width ratio of the voltage measuring contacts (L/W or sqr, the number of squares) on the Hall bar was 4.1 or 4.7 depending on which ohmic contact sets were used. The undamaged side of the Hall bar has the same geometry. Ohmic contacts consisted of Ti-Au or Au-Ge-Ni composition with no annealing needed for ohmic behavior at low temperatures. The magnetic easy axis is in-plane parallel to [1-10] on the GaAs (001) substrate with the uniaxial hard axis parallel to [110] GaAs (001), see Ref. [10, 16]. The out-of-plane direction [001], with respect to the underlying GaAs (001) substrate is a magnetic hard axis.

**C. FIB patterning**

The Ga$^+$ ion patterning equipment is an *FEI Helios$^{TM}$ Nanolab 650* system running at 30 keV. The devices were controllably damaged from wafer A after the mesa and ohmic contacts were fabricated. The calibrated dose levels were from a lower level of $1\times10^{13}$ Ga$^+$ ions/cm$^2$ to $\sim 1\times10^{16}$ ions/cm$^2$. Higher dose levels are available although at $1\times10^{16}$ ions/cm$^2$ the magnetic moment is already quenched and electrical measurements show no evidence of ferromagnetic properties; there is no AMR in $\rho_{xx}$ or anomalous Hall effect in $\rho_{xy}$ observable. There is possibly a small AMR signal below 0.03% although a random telegraph signal (RTS) noise pattern is developing at this high dose level and dominates the transport properties. This phenomenon also dominated the noise spectrum in Ga$^+$ FIB damaged InSb devices at the high dose levels [17], albeit in a different semiconductor material system with no capping layers. SRIM (Stopping Range of Ions in Matter), modelling [18] of the Ga$^+$ ion damage induced in MnSb shows non-uniform damage throughout the 50 nm thickness with the highest damage at the surface. Table [2] is a summary of the devices from wafer A that are reported in this paper.

| batch | device | damage dose (ions/cm$^2$) | FIB ion species |
|---|---|---|---|
| b(1) | c(1) to c(4) | undamaged | |
| b(2) | c(1) | $1\times10^{13}$ | Ga |
| b(2) | c(2) | $1.2\times10^{15}$ | Ga |
| b(2) | c(3) | $1\times10^{14}$ | Ga |
| b(2) | c(4) | $1.2\times10^{16}$ | Ga |

**Table [2].** A summary of the MnSb devices from wafer A.

The carrier density is not changing significantly with dose; although this is difficult to quantify via the normal Hall effect, although the conductivity (is sensitive to the dose level) and is decreasing with increasing $Ga^+$ ion dose. The conductivity does not reduce into the range (< $e^2/h$) with e the unit of charge and h Planck's constant, where a variable range hopping conductivity or a thermally activated transport would potentially dominate [17] or percolation effects would appear [11] although the effects of weak localization are present. Introducing defects and disorder into semiconductor devices with ion beam irradiation has been reviewed in Ref. [19] and the role that defects play in topological systems has also been investigated recently, [20] and is relevant to this work.

**D. Transport measurements**

Resistances were measured using constant AC currents of 500 nA at 33 Hz. Lock-in amplification was used with output offset nulling to improve the dynamic range, due to the generally small change in resistance signals with magnetic field. Resistance is converted to resistivity by dividing by (L/W) in the case of $\rho_{xx}$. Random noise levels in the $\rho_{xx}$ signal are typically 20 nV×$Hz^{-1/2}$, this is from a digitization limitation with the Stanford SR830 lock-in amplifier having a white noise frequency characteristic. The applied magnetic field range was either ±2 T for the Hall effect signal, or 8 T for magnetoresistance. The magnetic field was applied along the [001] direction of the GaAs substrate. Low temperature measurements were made between 1.5 and ~20 K typical range. This is below the reorientation transition temperature of the spins that has been reported in Ref. [2] as ~ 100 K and below the anomalous temperature dependences of the resistivity reported in Ref. [10].

**III. ELECTRICAL TRANSPORT**

**A. Resistivity temperature dependence**

The temperature dependence of the zero magnetic field resistivity ($\rho_o$) for temperature T < 100 K is very weak in the undamaged devices, i.e. $d\rho_o/dT \sim 0$, see Fig. 1.

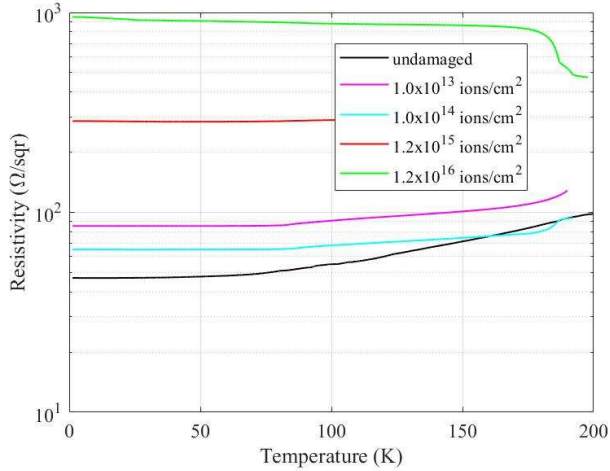

**FIG. 1.** Temperature dependent resistivity below ~ 200 K for an undamaged device and the Ga$^+$ ion FIB damaged devices.

At higher Ga$^+$ ion dose $d\rho_o/dT < 0$, and several signatures of localization appear. For example, on device b(2) c(2) from wafer A with a dose $1.2\times10^{15}$ ions/cm$^2$, $d\rho_o/dT < 0$ for temperatures below ~ 50 K. Then the change in conductivity with temperature, $\Delta\sigma_o \sim \frac{1}{\sqrt{2}} \times \frac{e^2}{h}$, in this region of $d\rho_o/dT < 0$ with $\sigma_o = 1/\rho_o$. At the higher dose levels, $d\rho_o/dT < 0$ from below 300 K. This conductivity change with temperature is an indication that weak localization and increasing disorder is becoming important at higher dose levels. There is a (small) variation in the resistivity between the undamaged devices, with $\rho_o$ varying from 30 to ~ 50 Ω/sqr in four nominally identical devices at 1.5 K.

## B. Magnetoresistance

The magnetoresistance has several components with the field (B) applied along the out-of-plane [001] direction. The low field is dominated by an anisotropic magnetoresistance effect relating to an EA in-plane switching, then a hard axis (HA) AMR signal. This saturates and is replaced by a log(B), characteristic of a weakly disordered two-dimensional system [21]. The amplitude of the low field AMR signal, depends on the level of damage and is linearly dependent on T, extrapolating to 0 at ~100 K. This signal is also hysteretic as expected even with B along [001]. The magnetic field is applied along [001], so the AMR is due to the HA out-of-plane response, however there is also an in-plane AMR effect and it is this that is hysteretic. There can also be a magneto-crystalline contribution [22] to this effect in MBE grown ferromagnetic materials although this is not contributing significantly here as the applied magnetic field is exclusively along the same crystal direction [001], in all measurements.

The GaOx cap preserves the surface states in the way that an Sb cap does, the AMR signal strength, $\Delta\rho_{xx}/\rho_{xx} = 0.120 \pm 0.005$ % at 1.5 K, comparable in size to previous devices [10]. This signal comes from the surface layer; this surface state has a quasi-2D (two-dimensional) behavior and this is discussed further in Sec. IV. The magnetoresistance signal is from the hard axis orientation B perpendicular to the (1-101) surface, but an in-plane component gives a hysteretic, easy-axis AMR contribution to $\rho_{xx}$.

Figure 2 shows the magnetoresistance to 8 T for undamaged device, b(1) c(2) from wafer A. The fit to a log(B) dependence is also shown by fitting $\rho_{xx}$ to log(B) from 1.7 T to 8 T. This procedure predicts the change in conductivity is from B = 0 T using the measured $\rho_{xx}$ at B = 0 T not a value in the fitting range of B where the intercept is determined.

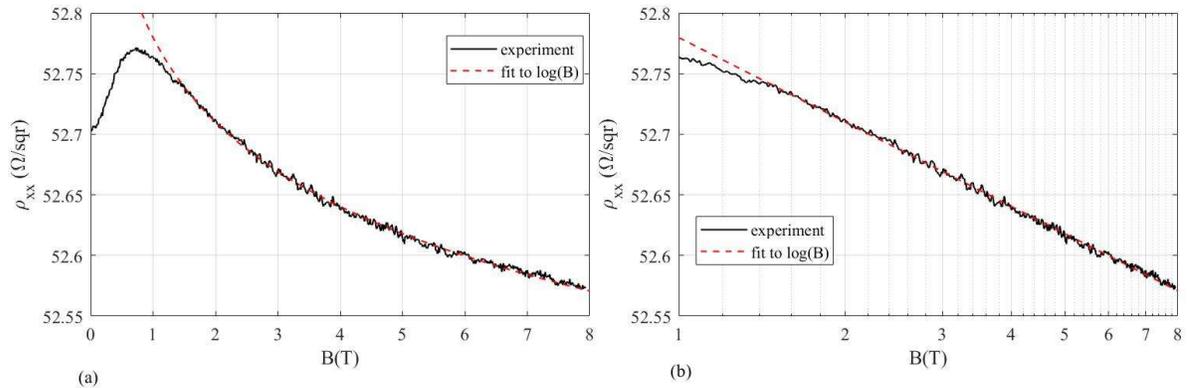

**FIG. 2.** Undamaged device, b(1) c(2) at 1.5 K showing (a) $\rho_{xx}$ on a linear B scale and (b) $\rho_{xx}$ on a log(B) scale. The fit to log(B) behavior is in the domain 1.7 to 8 T. Below ~1.7 T the AMR signal (positive magnetoresistance) due to easy-hard axis switching dominates.

In Fig. 2(b) the y-axis is plotted on a log scale to show the dependence on log(B). As the $Ga^+$ ion dose increases the dependence of $\rho_{xx}$ on magnetic field is still ~ log(B), however the fits become less convincing, see Fig. 3. Negative magnetoresistance behavior in disordered semiconductor systems has been attributed to wavefunction interference effects and this was reported in earlier work, Ref. [21, 23]. The results presented in this paper develop these arguments in a ferromagnetic device. The logarithmic dependence of $\rho_{xx}$ on B is also present in $\sigma_{xx}$ and plotting $\Delta\sigma_{xx}(B)$ enables a comparison between devices with different $Ga^+$ ion doses, as $\rho_o$ is changing significantly with dose. Figure 3 is a plot of $\Delta\sigma_{xx}(B)$ determined from the measured resistivities with respect to the conductivity at B = 0 T, i.e. $\Delta\sigma_{xx}(B) = \sigma_{xx}(B) - \frac{1}{\rho_o}$.

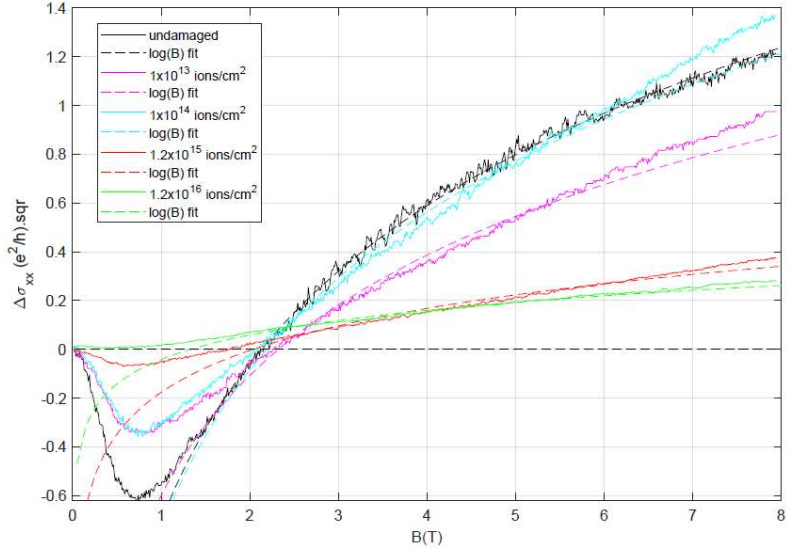

**FIG. 3.** $\Delta\sigma_{xx}$ as a function of magnetic field at 1.5 K for increasing Ga$^+$ ion dose levels up to ~$1.2\times10^{16}$ ions/cm$^2$. The dotted lines are least square fits to a log(B) dependence.

The y-axis units in Fig. 3 are scaled to the quantum conductance $e^2/h$. This fitting to log(B) is valid in the field region where the anisotropic magnetoresistance has saturated > 2 T and the AMR is no longer magnetic field dependent. The AMR signal is lost at high dose, see Fig. 3 for device b(1) c(4) at $1.2\times10^{16}$ ions/cm$^2$ at low applied magnetic field. At higher fields > ~ 2T, the magnetoconductance change in field follows a log(B) dependence. This is clear from the fits in Fig. 3. The dependence on B can be described by Eq. (1) and this has been seen in weakly disordered two-dimensional semiconductor devices [21].

$$\Delta\sigma_{xx}(B) = -\frac{e^2}{h} log_e(\frac{l_B}{l_e}) \qquad (1)$$

where $l_B$ is the magnetic length, given by: $\left(\frac{\hbar}{eB}\right)^{1/2}$, $l_e$ is the characteristic elastic scattering length and $\hbar$ is the reduced Planck's constant. This is assuming that the phase coherence length of holes in MnSb is greater than the film thickness, resulting in quasi-2D characteristics. The wavefunction interference that produces this contribution from Eq. (1) i.e. a negative magnetoresistance, is reduced at high magnetic field and the (larger) metallic conductivity is recovered; $\Delta\sigma_{xx}$ tends logarithmically to zero at high field. This is discussed further in Sec. IV. The results described in Ref. [21] is for a weakly disordered two-dimensional system and the MnSb devices reported here are a quasi-2D system with magnetic properties determined by a surface state [10] and with a level of disorder that is reflected in the magnetoresistance behavior beyond the saturation of the AMR effect at low magnetic field.

**C. Anomalous Hall effect**

The measured $\rho_{xy}$ signal can contain an unwanted $\rho_{xx}$ contribution (due to the effects of disorder or processing non-uniformity), and this can be removed by symmetrizing the Hall data, via Eq. (2). This limits further analysis of the Hall signal to one sense of magnetic field and any hysteretic signal is lost by this procedure but it does remove the (significantly larger) $\rho_{xx}$ signal interference.

$$\rho_{xy} = \tfrac{1}{2} \times [\rho_{xy}(+B) - \rho_{xy}(-B)] \qquad (2)$$

In Fig. 4 $\rho_{xy}$ is plotted for wafer A devices up to 2 T at 1.5 K after using Eq. (2) to remove $\rho_{xx}$ components, as discussed. The Hall voltage is negative as the experimental set up is calibrated for electron-based structures that have a defined positive Hall voltage for defined positive magnetic field sense. This comparison confirms the p-type behavior in the MnSb devices. There are (at least) two contributions to $\rho_{xy}$ and the expression for $\rho_{xy}$ is given by Eq. (3) when there are two different contributions to the Hall signal. The first term in Eq. (3) is the normal Hall effect (with Hall constant $R_H$) and the second term is the anomalous Hall effect contribution, with anomalous Hall constant $R_{AH}$.

$$\rho_{xy}(B) = R_H \times B + R_{AH} \times M(B) \qquad (3)$$

$M(B)$ is the magnetization of the MnSb. The anomalous Hall effect term saturates at magnetic field $B_{sat}$, typically ~ 1 T in the undamaged material. This switch over to the normal Hall effect with increasing field can be seen in Fig. 4.

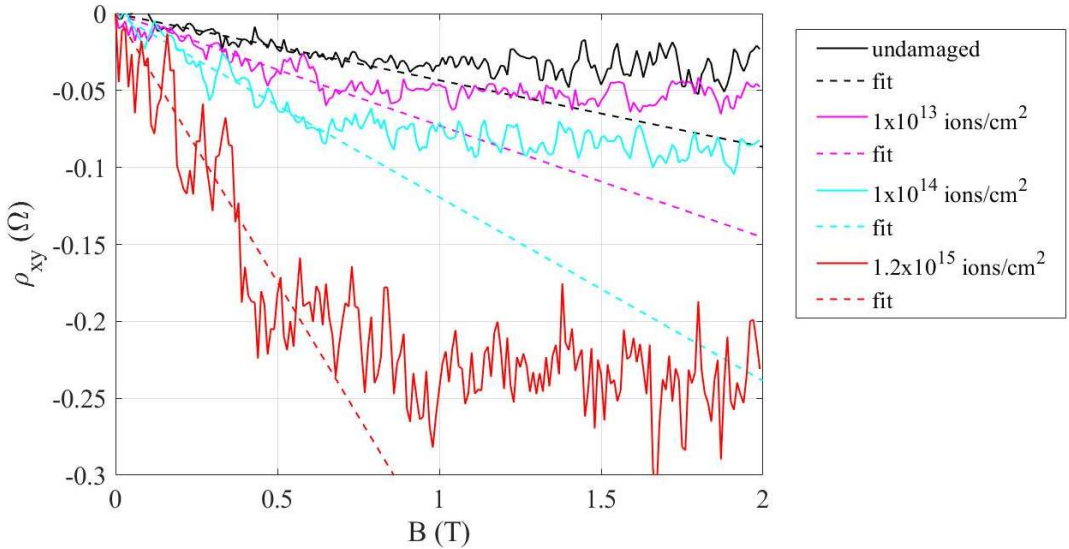

**FIG. 4.** $\rho_{xy}$ as a function of magnetic field for different Ga$^+$ ion dose levels at 1.5 K. There is no AHE at the highest Ga$^+$ ion dose level, so the data for device b(2) c(4) is not shown. The fits are to a linear dependence to determine $R_{AH}$ below $B_{sat}$. The increasing noise level with dose is already apparent in $\rho_{xy}$ at $1.2\times10^{15}$ ions/cm$^2$.

Although the noise level looks high in this data plot, it is rather that the signal size is small. Coadding of multiple data sets is used to reduce the random noise level, for example in Fig. 4 four field sweeps are averaged for each dose level, in addition to using Eq. (2) to reduce the influence of $\rho_{xx}$, so each data point is the mean of eight independent measurements. FIB damaged wafer A device b(2) c(4) with a dose $1.2\times10^{16}$ ions/cm$^2$ does not show a clear, reproducible anomalous Hall effect signal, it is dominated by larger RTS noise. This noise signal in $\rho_{xy}$ is greater than the characteristic 20 nV×Hz$^{-1/2}$ noise signal present in $\rho_{xx}$.

The anomalous Hall resistance is dependent on the particular scattering mechanisms that contribute to resistivity $\rho_o$, with $\rho_{xy} \sim \rho_o^\beta$, see Ref. [24, 25]. The power coefficient $\beta$ can be determined experimentally by varying the measurement temperature or an alternative way to estimate $\beta$ is to change the impurity concentration, hence $\rho_o$ at constant temperature or to FIB damage the material. The undamaged MnSb devices are in the high conductivity regime ($k_f.l_e \gg 1$) where $k_f$ is the Fermi wavevector and $l_e$ is the elastic scattering length defined earlier. In the damaged devices this criterion is still valid although $k_f.l_e$ reduces with ion dose. In this high conductivity regime, skew scattering is the dominant contribution to the anomalous Hall effect [25], with $\beta \approx 1$. This behavior is evident in Fig. 5, where the anomalous Hall constant is plotted as a function of $\rho_o$ for the damaged and undamaged devices.

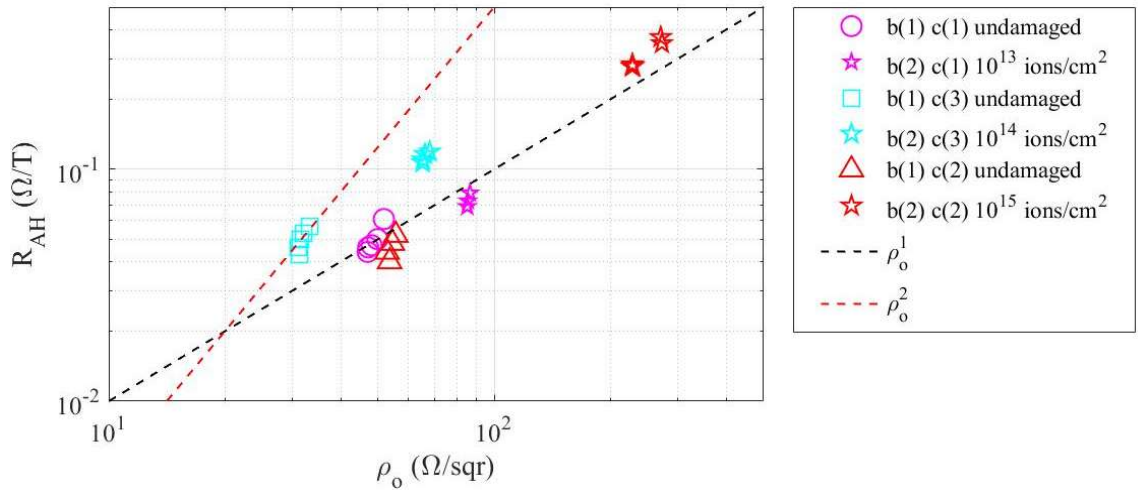

**FIG. 5.** The anomalous Hall effect constant as a function of resistivity as FIB damage is introduced.

The black dashed line in Fig. 5 is the $\beta = 1$ generic dependence that is characteristic of skew scattering, an impurity scattering mechanism driven by spin-orbit interaction. The experimental AHE signal scales with $\rho_o^1$ as the disorder level increases but for a fixed level of disorder the AHE signal tends to vary as $\sim \rho_o^2$ at finite temperature (red dashed line), not with $\beta = 1$. This is shown in Fig. 5 as the cluster of data points are plotted from 1.5 K to ~ 20 K for each dose level, albeit they have the same-colored symbols. The temperature dependences are discussed in more detail in Sec. IV. An inelastic phonon scattering contribution can increase $\beta$ from 1 to 2 as is broadly evident here. A second possible scattering mechanism is side-jump scattering [25] and this effect changes the relationship to: $\rho_{xy} \sim \rho_o^2$, although a skew scattering is dominant. At the highest $Ga^+$ ion dose level the AHE is lost from $\rho_{xy}$ with RTS dominating this signal. Other mechanisms can contribute to AHE signals, see Ref. [26] for example, although they are unlikely to be present in MnSb.

**IV. DISCUSSION**

The $Ga^+$ FIB damaged devices have dose levels from $1\times10^{13}$ to $1.2\times10^{16}$ ions/cm$^2$, see Table (1). FIB damage reduces the AMR but increases the AHE constant until the AHE signal is lost at a dose level of ~ $1.2\times10^{16}$ ions/cm$^2$. The transport properties in the ferromagnetic regime are in general agreement with previous detailed transport measurements comparing Sb-terminated and non-terminated MnSb devices at low temperature [10]. The AHE was stronger in the non-Sb surface terminated MnSb devices (transport in that case is dominated by bulk states) with generally higher scattering rates present in this device material. The fits of $\Delta\sigma_{xx}(B)$ to a log(B) behavior is expected [21] and deviations from this in the damaged devices, see Fig. 3. are possibly due to current shunting in the MnSb closer to the GaAs interface or non-uniform FIB damage along the Hall bar. SRIM modelling [18] of the damage profile indicates that the surface region is predominantly affected in the case of $Ga^+$ ion patterning.

The temperature dependence of the transport coefficients in MnSb is complicated by the existence of a previously reported anomalous dependence of $\rho_{xx}$ and $\rho_{xy}$. Previously reported work on MBE grown GaSb/Mn digital alloys, see Ref. [27] showed a temperature dependent negative magnetoresistance with the possible existence of multiple ferromagnetic phases, some of which could be comparable to $Ga^+$ ion damaged MnSb. This behavior would perhaps be observable in the log(B) dependent $\rho_{xx}$ at higher fields, although this was not reported in Ref. [27] for comparison. The AHE signal strengthens with increasing temperature, whereas the AMR signal size reduces at higher temperature at least until 40 K. Figure 6(a) shows $R_{AH}$ as a function of temperature up to ~ 20 K. The variation in the (red) data points at the highest dose is due to the data set consisting of two cool down measurements. This gives an approximate error in the $R_{AH}$ signal determination of ±12.5 % on the mean value of 0.31 $\Omega$/T for the highest level of damage before the AHE signal is lost into RTS noise at a dose level of $1.2\times10^{16}$ ions/cm$^2$. This effect of increased noise characteristic was seen previously in MnSb over a broader temperature range.

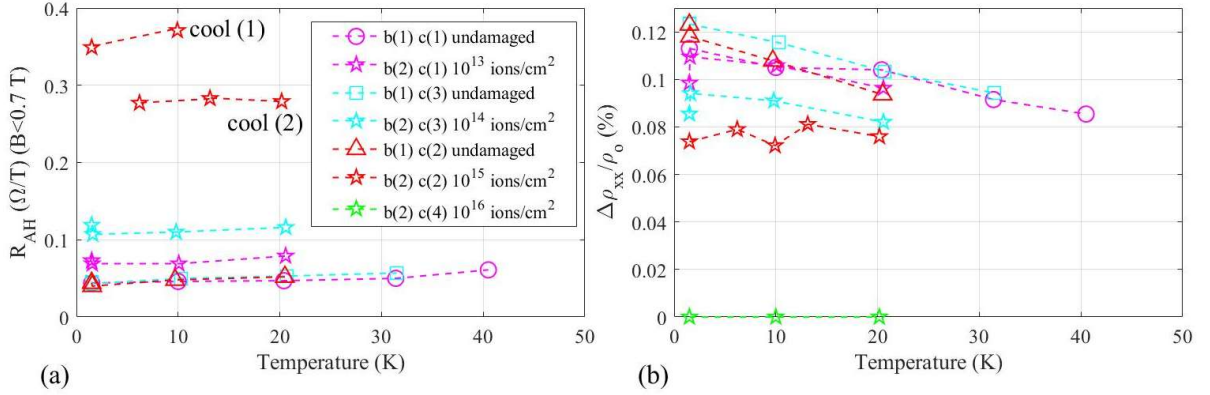

**FIG. 6.** (a) The anomalous Hall constant for different dose levels as a function of temperature, (b) The resistivity change due to AMR as a function of temperature. Experimental data from two separate cool downs is shown for device b(2) c(2).

In Fig. 6(b) the AMR signal reduces with increasing temperature at least until ~ 40 K and extrapolates (approximately independent of dose level) to 0 at ~ 100 K, consistent with the temperature measured in Ref. [2] where the sign of the AMR signal changes above ~ 100 K, although this has not been investigated in this paper. At low temperature $\Delta\rho/\rho_o$ is ~ 0.120 ± 0.005 % in the undamaged MnSb devices with this signal reducing to zero at the highest Ga$^+$ ion dose, indicating that the ferromagnetic signal has been quenched in the device. There are no transport measurements that indicate a transition to thermally activated conductivity that would be characteristic of an amorphous device at the highest dose level.

Fully disordered metallic characteristics are still not present even at the highest Ga$^+$ ion dose levels where the expectation is that $\rho_{xy} \sim \rho_o^\beta$, where $\beta > 1$ in the disordered metal regime. A parameter that is more sensitive to disorder is the anomalous Hall conductivity, see Eq. (4) below:

$$\sigma_{xy} = \frac{\rho_{xy}}{(\rho_{xx}^2+\rho_{xy}^2)} \sim \frac{R_{AH} \times B_{sat}}{\rho_{xx}^2} \qquad (4)$$

with $\rho_{xx} \gg \rho_{xy}$. $\sigma_{xy}$ from the anomalous Hall effect reduces in size with dose rather than the increasing signal of $\rho_{xy}$. $B_{sat}$ is the saturation field, $\mu_o \times M_{sat}$ (with $\mu_o \approx 4\pi\times10^{-7}$ H/m; $M_{sat}$ the saturation magnetization) where the AHE contribution to $\sigma_{xy}$ is then a constant quantity with only the normal Hall effect contributing. Figure 7 shows the anomalous Hall constant and the anomalous Hall conductivity determined from Eq. (4) as a function of the AMR relative resistance change size. The quasi-2D nature of the surface states reported here means that the quantity $\sigma_{xy}$ does not have a unit of length associated with it, as it would for a three-dimensional system. The normal Hall coefficient is so small that the only contribution to $\sigma_{xy}$ is from the AHE and this saturates above $B_{sat}$ which is ~ 1 T with $B_{sat}$ reducing with dose level. Figure 7(a) shows $R_{AH}$ which increases *both* with the damage level and the increasing scattering rate.

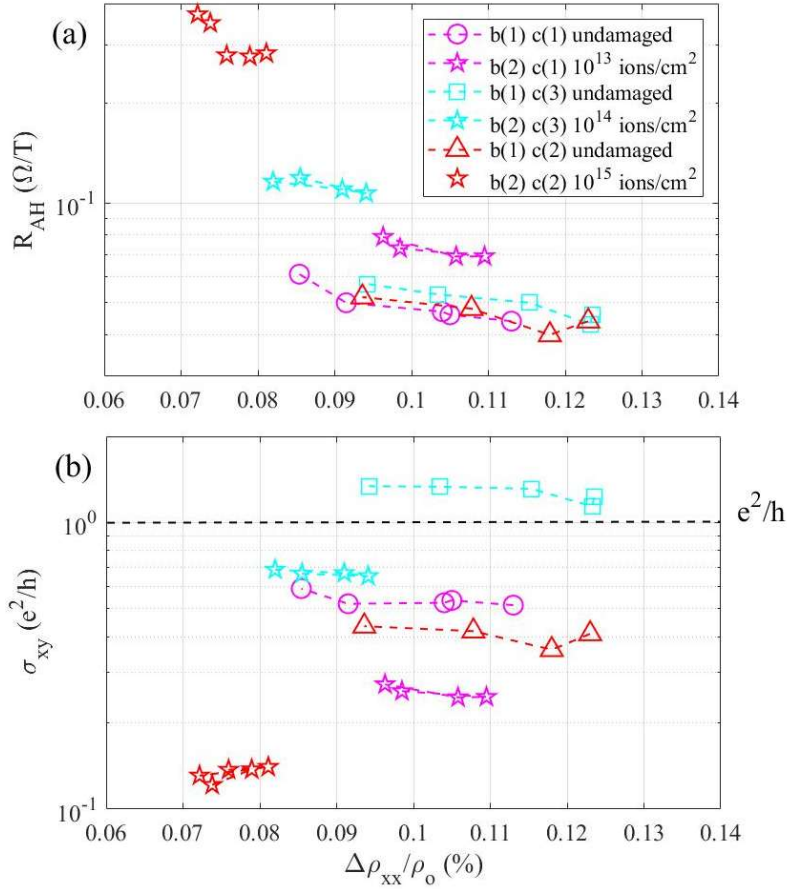

**FIG. 7.** Summary of parameters from the anomalous Hall effect for the FIB damaged and undamaged devices. (a) $R_{AH}$ as a function of the size of the AMR signal and (b) $\sigma_{xy}$ as a function of the size of the AMR signal. A quantum anomalous Hall effect signal would have a universal $\sigma_{xy}$ contribution at $e^2/h$ conductivity.

In Fig. 7(b) $\sigma_{xy}$ is plotted as a function of AMR relative resistance change size. In the undamaged devices $\sigma_{xy} \sim$ (0.4 to 1.2)$\times e^2/h$. This range of $\sigma_{xy}$ values arises primarily from by the range of $\rho_o$ values between the different Hall bars. Although $\sigma_{xy}$ is close to $e^2/h$, which is the universal value for $\sigma_{xy}$ in the quantum anomalous Hall effect [25] this is coincidental here as no inverted bands or topological states are expected to be stable for any surface states in the niccolite phase of MnSb. In the anomalous Hall effect there is a further regime, that of the intrinsic or scattering-independent behavior with $\rho_{xy} \neq \rho_o^\beta$. This is the Berry phase contribution region [Jungwirth] and already reported experimentally in Ref. [2] for the case of devices in MnSb.

## V. CONCLUSIONS

Focused Ion Beam damage of ferromagnetic MnSb:GaAs(001) devices has been demonstrated. The resulting properties of the electrical resistivity $\rho_{xx}$ and $\rho_{xy}$ with increasing ion-beam dose has led to an understanding of how the ferromagnetic properties could in principle be modulated on a submicron length scale. A maximum $Ga^+$ ion dose of $\sim 10^{16}$ ions/cm$^2$ is able to fully quench the ferromagnetic behavior, with no measurable anisotropic magnetoresistance in $\rho_{xx}$ at low applied magnetic field and no anomalous Hall effect signal in $\rho_{xy}$, indicating no electrical evidence of ferromagnetic properties. The conductivity $\sigma_{xx}$ has a log(B) dependence in a magnetic field for all $Ga^+$ ion dose levels and this is characteristic of a weakly disorder metal with conductivity change $\sim e^2/h$ at high magnetic field. There is a reduction in the saturation magnetization with $Ga^+$ ion dose and this is apparent in the AHE. The AHE contribution to $\sigma_{xy}$ in MnSb is also $\sim e^2/h$ and is then sensitive to the level of disorder introduced during the FIB process, reducing with dose level and is a useful parameter for assessing residual magnetic behavior in the devices. GaSb and GaOx have both been used as capping layers for these devices and both are compatible with preserving the surface states from oxidation. $Ga^+$ ion damage at 30 keV is predominantly in the surface region and this has advantages for materials like MnSb that have surface properties distinct from bulk states with crystalline integrity maintained into the paramagnetic regime. This is an important result that is supported by both simulations and experiments. This would allow *in-situ* patterning of MBE-grown ferromagnetic materials followed by regrowth of additional capping layers, altermagnetic layers or barrier layers for two-dimensional hole confinement or magnetic exchange coupling schemes. Patterning surface states of Topological Insulator devices would also benefit from this type of processing and would be an exciting development.

**Data availability statement**

The data that support the findings of this study are openly available at the following URL/DOI: https//doi.org/zenodo

**Acknowledgements**

This research was supported at University College London by an EPSRC (UK) Programme Grant, Number EP/R029075/1, Non-Ergodic Quantum Manipulation. Technical support for the FIB patterning was provided by E. W. Tapley at the Cavendish Laboratory, University of Cambridge, UK.

**Conflict of interest**

The authors have no conflicts of interest to disclose.

**E-mail addresses:**

stuart.holmes@ucl.ac.uk,

jonathan.gough.18@ucl.ac.uk

ethan.dommett@warwick.ac.uk

gavin.bell@warwick.ac.uk

**ORCID iDs**

S. N. Holmes https://orcid.org/0000-0002-3221-5124

J. Gough https://orcid.org/0009-0003-9951-7658

E. Dommett https://orcid.org/0009-0008-2897-9800

G. R. Bell https://orcid.org/0000-0002-6687-7660